\PassOptionsToPackage{table,xcdraw,svgnames}{xcolor}
\PassOptionsToPackage{colorlinks,pagebackref,bookmarks}{hyperref}

\documentclass[preprint,journal]{vgtc}               

\graphicspath{{figures/}{pictures/}{images/}{./}} 

\usepackage{times}                     

\usepackage{tabu}                      
\usepackage{booktabs}                  
\usepackage{lipsum}                    
\usepackage{mwe}                       

\usepackage{mathptmx}                  

\usepackage{amssymb}
\usepackage{amsmath}
\usepackage{multirow}
\usepackage{tcolorbox}
\tcbuselibrary{breakable}
\usepackage{soul}
\usepackage{array}
\usepackage{xspace}
\usepackage{caption}
\usepackage{multirow}
\usepackage{ragged2e}
\usepackage{balance}
\usepackage{enumitem}
\usepackage{subcaption}
\usepackage{float}
\usepackage{stfloats}

\usepackage{hyperref}

\onlineid{1274}

\vgtccategory{Research}

\vgtcinsertpkg

\preprinttext{To appear as a short paper at IEEE VIS 2024.}

\newcommand{\sqboxs}{\fontcharht\font`X}
\newcommand{\sqbox}[1]{\textcolor{#1}{\rule{\sqboxs}{\sqboxs}}}
\newcommand{\sqboxblack}[1]{\setlength{\fboxsep}{0pt}\fbox{\sqbox{#1}}\xspace}

\definecolor{propDarkOrange}{HTML}{E6550D}
\definecolor{propLightOrange}{HTML}{FDAE6B}
\definecolor{propDarkGray}{HTML}{636363}
\definecolor{propLightGray}{HTML}{bdbdbd}

\definecolor{cdfDarkOrange}{HTML}{D94701}
\definecolor{cdfLightOrange}{HTML}{FD8D3C}
\definecolor{cdfDarkGray}{HTML}{000000}
\definecolor{cdfLightGray}{HTML}{969696}

\definecolor{yaOrange}{HTML}{F28E2B}
\definecolor{plaBlack}{HTML}{000000}

\definecolor{ncFG}{HTML}{4589FF}
\definecolor{pcBG}{HTML}{F4F4F4}
\definecolor{pcFG}{HTML}{002D9C}
\definecolor{ncBG}{HTML}{262626}

\newcommand{\iDotE}{i.\,e.,\xspace}
\newcommand{\eDotG}{e.\,g.,\xspace}

\newcommand{\meanicon}{\raisebox{-.8\dp\strutbox}{\includegraphics[height=2.35ex]{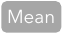}}\xspace}

\title{Dark Mode or Light Mode? Exploring the Impact of Contrast Polarity on Visualization Performance Between Age Groups}

\author{Zack While\thanks{e-mail: zwhile@cs.umass.edu} %
\and Ali Sarvghad \thanks{e-mail: asarv@cs.umass.edu}}
\affiliation{\scriptsize University of Massachusetts Amherst}

\abstract{
This study examines the impact of positive and negative contrast polarities (\iDotE light and dark modes) on the performance of younger adults and people in their late adulthood (PLA). In a crowdsourced study with 134 participants (69 below age 60, 66 aged 60 and above), we assessed their accuracy and time performing analysis tasks across three common visualization types (Bar, Line, Scatterplot) and two contrast polarities (positive and negative). We observed that, across both age groups, the polarity that led to better performance and the resulting amount of improvement varied on an individual basis, with each polarity benefiting comparable proportions of participants. However, the contrast polarity that led to better performance did not always match their preferred polarity. Additionally, we observed that the choice of contrast polarity can have an impact on time similar to that of the choice of visualization type, resulting in an average percent difference of around 36\%. These findings indicate that, overall, the effects of contrast polarity on visual analysis performance do not noticeably change with age. Furthermore, they underscore the importance of making visualizations available in both contrast polarities to better-support a broad audience with differing needs. Supplementary materials for this work can be found at \url{https://osf.io/539a4/}.
}

\begin{document}


\firstsection{Introduction}

\maketitle

\firstsection{Introduction}

The visual analysis performance of \textbf{P}eople in the \textbf{L}ate \textbf{A}dulthood (PLA) development stage~\cite{berk2022development} may differ than those in early and middle stages of adulthood development~\cite{while2024gerontovis}. For instance, in a recent user study, While et al.~\cite{while2024glanceable} found that participants 65 and older took notably more time than younger participants in performing a basic data comparison task across certain combinations of glanceable visualizations designs and data sizes on a smartwatch. Prior work in aging and information design suggests several factors---including age-related physiological changes, generational gaps in technology literacy, and socio-economic differences---could contribute to varying performance among individuals at different stages of adult development~\cite{alexander2011passive, johnson2017designing, while2024gerontovis}.

In this study, we investigated how visualization contrast polarity design may impact younger adults' (YA) and PLA's performance. Polarity is commonly divided into \textit{positive} and \textit{negative} contrasts. Positive contrast polarity, also known as \textit{light mode}, involves dark foreground objects (\eDotG black text) on a light background (\eDotG white.) Negative contrast polarity, also known as \textit{dark mode}, involves light foreground objects (\eDotG white or light gray text) on a dark background (\eDotG black or dark gray). 

Contrast polarity is a \textit{non-data-encoding} visualization design choice, which means it does not pertain to the process of selecting effective graphical marks (\eDotG area) and perceptual channels (\eDotG shape) for representing data values. Nonetheless, non-data-encoding visualization design choices can significantly interact with people's ability to effectively use visualizations. For instance, small axis labels and low contrast can hinder PLA's ability to use visualizations~\cite{morey-heart-failure, whitlock}. Outside of the context of data visualization, vision research has also shown that positive contrast designs can improve PLA's accuracy, speed, and endurance of reading text~\cite{piepenbrock2014positive,piepenbrock2013positive,sethi2023dark}. Related to visualization, negative contrast may enhance performance by reducing glare and brightness, eye strain, and fatigue. On the other hand, it may hinder performance by diminishing the clarity and perceptibility of some graphical details~\cite{johnson2017designing}. However, we do not know if such conjectures would translate to tangible and measurable differences in PLA's performance. To fill this knowledge gap, we conducted a crowdsourced study on Prolific~\cite{Prolific80:online} involving 69 participants below the age of 60 and 66 participants aged 60 and above in which we explored how contrast polarity impacts visual analysis performance as well as whether these impacts change with age. We assessed participants' accuracy and time using three common visualizations: Bar, Line, Scatterplot, each depicted using positive and negative contrast polarity. 

Overall, we did not find a consistent difference in performance along the axes of age and contrast polarity. In each age group, there were participants who performed better with each contrast polarity, however the proportions of each polarity were similar between and within age groups. This suggests that perhaps no universal recommendations regarding an ``optimal'' contrast polarity can be reliably made. However, the differences in individual performance indicate the importance of creating visualizations that support both polarities and incorporating  mechanisms that enable individuals to customize polarity according to their personal preferences. 

\section{Related Work}
This section discusses topics related to contrast polarity in the context of PLA and data visualization.
Contrast polarity is the degree of contrast between two visual elements, categorized as \textit{negative} (light foreground objects on dark background) or \textit{positive} (dark foreground objects on light background). This concept has become more prominent in user interface design and user preferences due to recent increases in support of both light and dark modes by website and app developers ~\cite{andrew2024light}.
Although there are occasionally inverse versions of these definitions in other existing work, we will use the aforementioned definition throughout this paper to avoid confusion.

While research at the intersection of data visualization and contrast polarity is limited, vision studies have explored its effects on reading speed and accuracy (\eDotG~\cite{chung2009contrast}), perceptual closure (\eDotG~\cite{spehar2002role}), visual acuity (\eDotG~\cite{westheimer2003visual}), and low vision (\eDotG~\cite{rubin1989psychophysics}). Piepenbrock et al.~\cite{piepenbrock2013positive} found that positive contrast designs improved visual acuity and proofreading for both older and younger adults, suggesting that increased ambient light might enhance visual capabilities without sacrificing perceived contrast. Notably, their study differed from ours in task type, setting, and participant demographics. Our study was crowdsourced and involved visual analysis tasks on computer screens, while they focused on identifying the orientations of Landolt C optotypes in a dark laboratory with controlled lighting. Sethi and Ziat~\cite{sethi2023dark} concluded that older adults' preference for positive contrast was based on minimizing mental fatigue while younger adults preferred negative contrast based on aesthetics. Outside academia, contrast design standards primarily stem from organizations like the World Wide Web Consortium~\cite{wcag2023}, offering guidelines generally applicable to data visualization despite focusing on text and color blocks. Industry vendors have also provided contrast design recommendations for app interfaces~\cite{ibm-carbon-design, apple-visual-design, google-material-design}, although their effects on PLA's performance remain unexplored.

\section{Study Design}
\subsection{Stimuli}
We utilized a version of the software created for an online study by Saket et al.~\cite{saket2018task} that was graciously provided by the original authors, making a small number of modifications to fit our study. First, we only used the Bar Chart, Line Chart, and Scatterplot visualizations to keep the scope of the study manageable.  For similar reasons, we then reduced the set of tasks to two options: \textit{Find Clusters}, where participants counted the number of groups depicted in a visualization, and \textit{Characterize Distribution}, where participants described the spread of the data in a visualization. This resulted in a total of 24 stimuli per task, based on 4 different data subsets and questions per combination of the 3 visualization types and 2 contrast polarities. Beyond the changes stated here, the content of the stimuli otherwise matched that of the original software~\cite{saket2018task}.

The color palettes for both contrast polarities were from IBM's \textit{Carbon Design System}~\cite{ibm-carbon-design}; the hexadecimal value of background hues in positive and negative contrast polarity were \#F4F4F4~\sqboxblack{pcBG} and \#262626~\sqboxblack{ncBG}, respectively, while the foreground hues were \#002D9C~\sqboxblack{pcFG} and \#4589FF~\sqboxblack{ncFG}, respectively.
We also checked that the stimuli followed existing digital design guidelines for PLA, such as a minimum contrast ratio of 4.5:1 between foreground and background objects as per Web Content Accessibility Guidelines (WCAG)~\cite{Developi34:online}.
\autoref{fig:viz-samples} provides examples of our study stimuli. 

\definecolor{exp-bg}{RGB}{22,22,22} 
\begin{figure*}[t]
\centering
\subfloat[Positive, Bar Chart]{
\centering
\fbox{\includegraphics[height=0.082\textwidth]{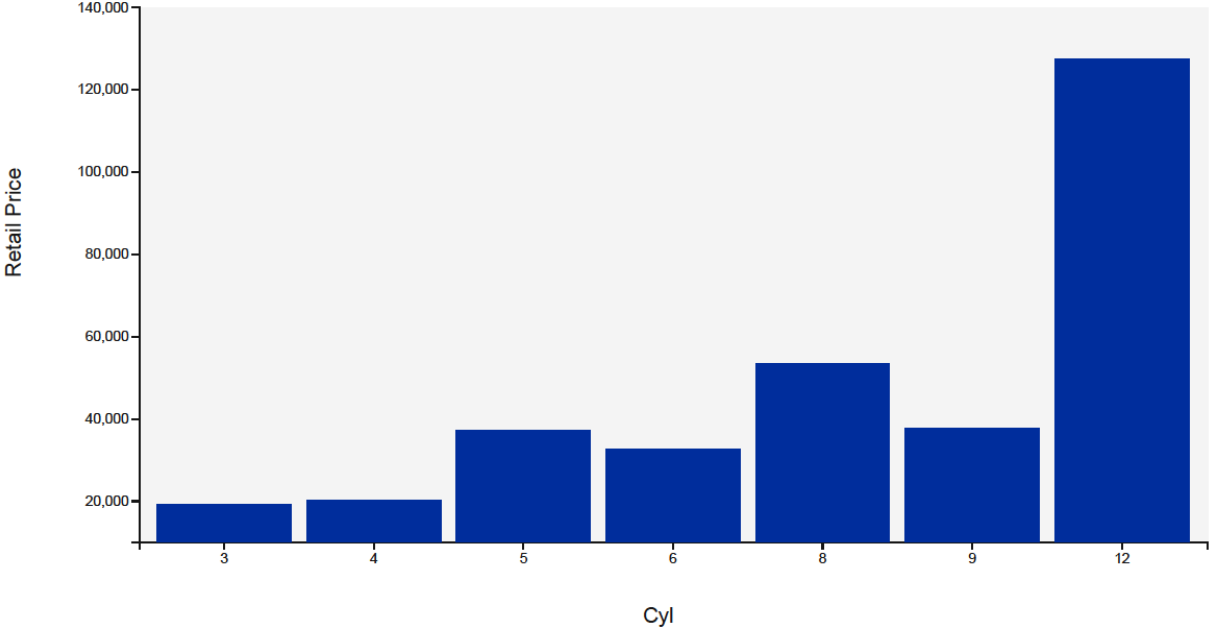}}
}
\subfloat[Positive, Line Chart]{
\centering
\fbox{\includegraphics[height=0.082\textwidth]{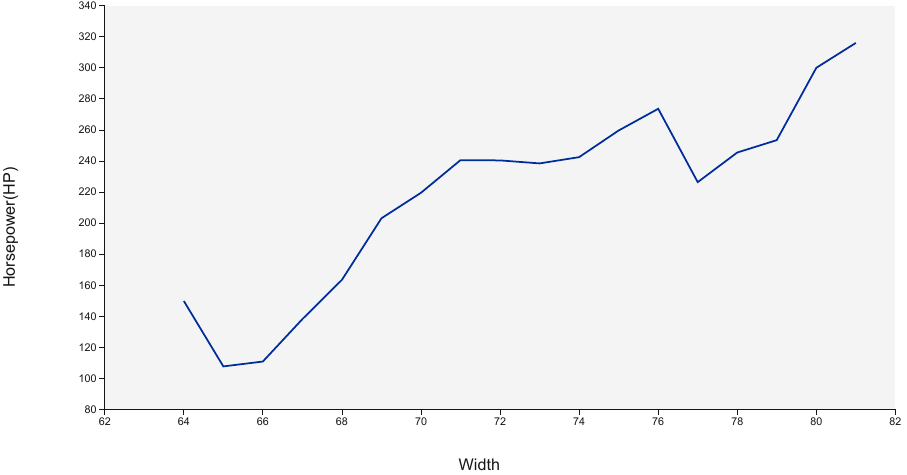}}
}
\subfloat[Positive, Scatterplot]{
\centering
\fbox{\includegraphics[height=0.082\textwidth]{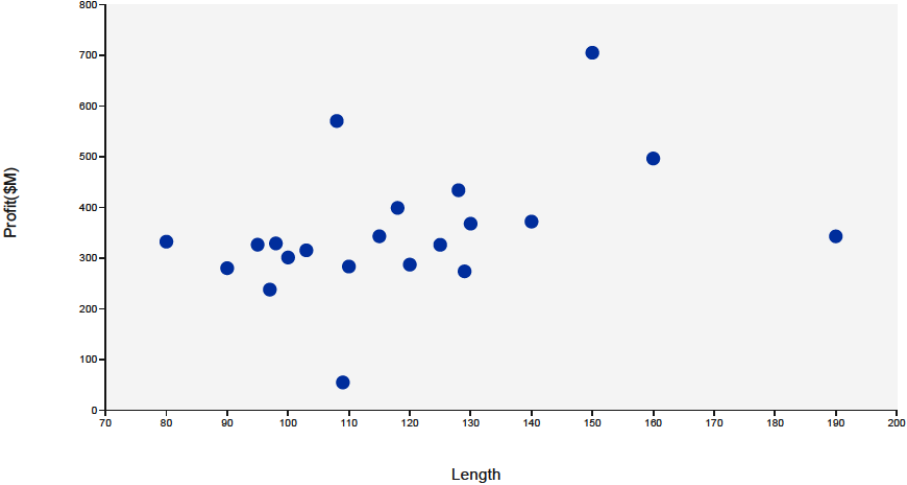}}
}
\subfloat[Negative, Bar Chart]{
\centering
\colorbox{exp-bg}{\includegraphics[height=0.082\textwidth]{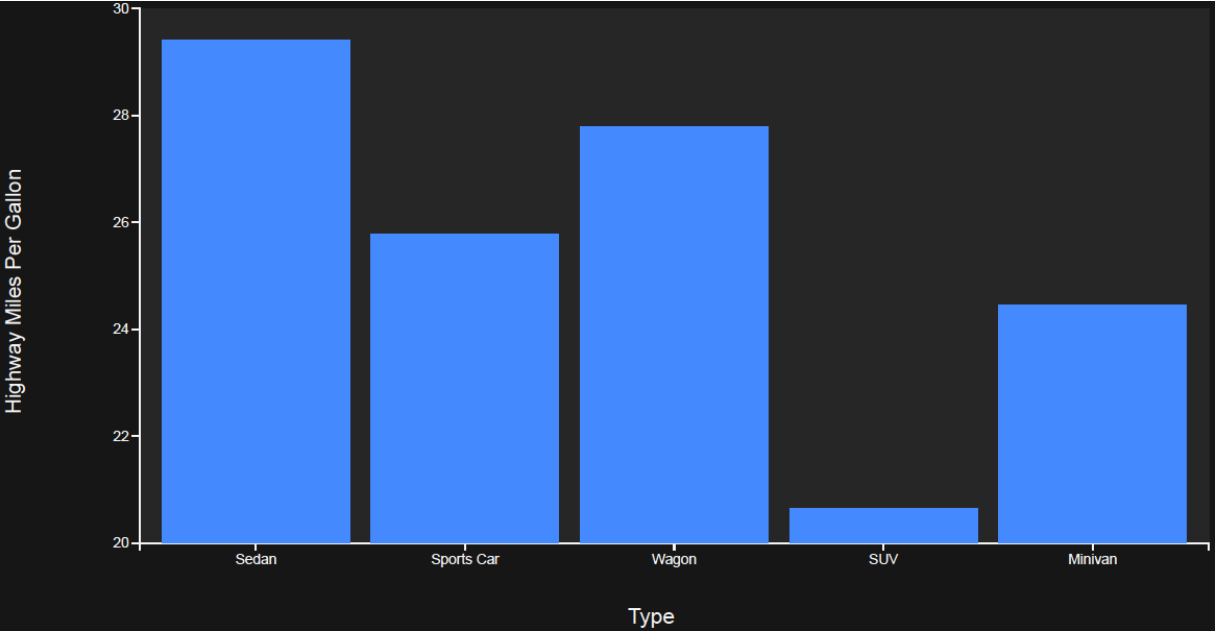}}
}
\subfloat[Negative, Line Chart]{
\centering
\colorbox{exp-bg}{\includegraphics[height=0.082\textwidth]{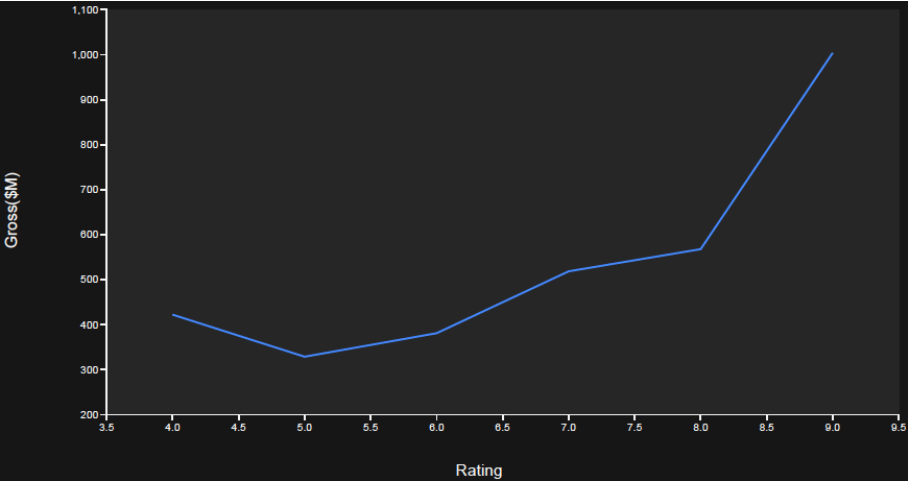}}
}
\subfloat[Negative, Scatterplot]{
\centering
\colorbox{exp-bg}{\includegraphics[height=0.082\textwidth]{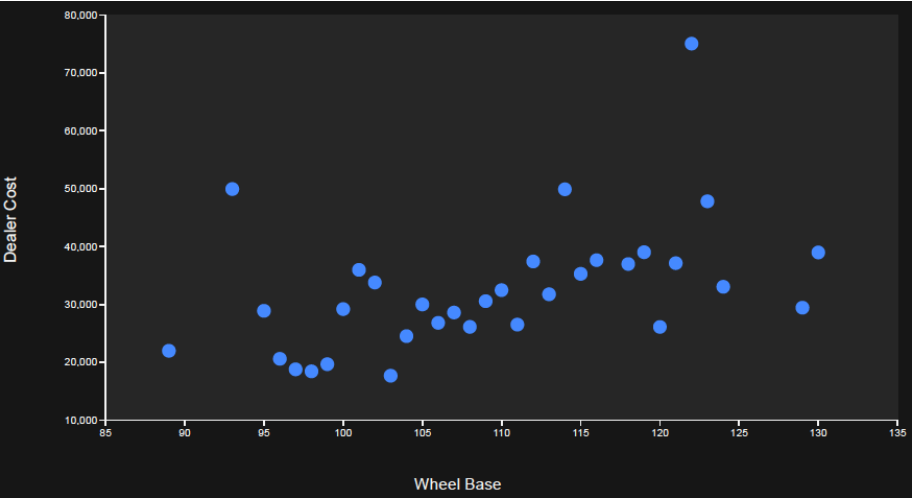}}
}
\caption{Example study stimuli, combinations of visualization (Bar Chart, Line Chart, Scatterplot) and contrast polarity (positive, negative).}
\label{fig:viz-samples}
\end{figure*}

\subsection{Participants and Procedure}
We used the Prolific \cite{Prolific80:online} crowdsourcing platform to recruit 135 US-based participants; 69 were YA (below age 60) and 66 were PLA (60 and older). All participants were required to have an approval rate of 98\% or higher and to use a desktop computer to maintain a more consistent screen size. Participants started the study by filling out a questionnaire with basic demographic data (\eDotG age, gender). We also asked about aspects such as education level, visualization familiarity, visions condition (\eDotG colorblindness), preferred contrast polarity on electronic devices, use of electronic devices to track activity and health, and use of any accessibility tools (\eDotG screen readers). We only observed a small impact on performance by visualization familiarity, otherwise finding no impact by demographic factors (more details in the supplemental materials).
Next, each participant completed 24 trials. Each trial included a visualization accompanied by a four-alternative forced-choice (4AFC) question. 
The order of trials was randomized and counterbalanced, and we collected participants' accuracy (correct/incorrect) as well performance time (seconds) per trial. The study took participants between 20 and 40 minutes, and each was paid \$13.

\section{Data Analysis and Results}\label{sec:study-2}

To begin our data analysis, we first evaluated participants' work quality by checking their score on an easy-to-answer ``test" question to ensure they were not replying to questions randomly, removing any participants who failed the test. These questions, provided about halfway through the experiment, would (1) introduce and show a data visualization, presenting the user with only two options to select (\textit{True} or \textit{False}) and (2) directly ask the user to select \textit{True}. Incorporating such test questions with obvious answers in crowdsourced studies is a common approach for flagging possible random performance and is consistent with the original software~\cite{saket2018task}. In total, we removed 8 YA and 7 PLA who failed this test. 
We also filtered out 9 YA and 7 PLA who indicated that they used accessibility tools/features on their computers, as we could not control for their impact on performance. This resulted in a final participant pool of 52 YA and 52 PLA. Lastly, we used a Hampel filter to remove any trials with times more than 3 median absolute deviations from the median for a given age group~\cite{pearson1999data}, as we observed some very large outliers (some up to 10 minutes) that we could not reasonably explain. In the rest of this section we present the findings of our data analysis, organized by accuracy and time; note that we observed similar patterns for both the \textit{Find Clusters} and \textit{Characterize Distribution} tasks~(more details available in the supplemental materials), so we have merged their results. 

\subsection{Initial Analysis Method}
A preliminary analysis of participants' mean performance (available in the supplemental materials) indicated a lack of evidence for contrast polarity having any effect on performance for either age group at both the overall and per-visualization levels. However, it appeared plausible that this result was partly due to an underlying structure of some participants performing well with one contrast polarity and poorly with another, resulting in their overall performance ``evening out'' when calculating the bootstrapped means and confidence intervals. 
Motivated by this observation as well as recent work emphasizing the impact of individual differences on performance~\cite{davis2022risks}, we focused our analysis at the individual level to see if contrast polarity was more impactful for some participants than others and if the impact was consistent with age. 

\subsection{Further Analysis Methods}
To investigate how contrast polarity may impact the ability to perform visual analysis tasks, we calculated each participant's mean performance across all trials for each of the contrast polarities. We then calculated the \textit{percent difference} between their two averages for a given metric (\eDotG accuracy), which we will define as the ratio between two measurements' absolute difference and their mean. In other words, for measurements $m_1$ and $m_2$ the percent difference would be 
$\frac{|m_1-m_2|}{\tfrac{1}{2}(m_1+m_2)}$. 
This method is more robust than signed differences, as it is scaled per participant, symmetric (\iDotE order of $m_1$ and $m_2$ does not matter), and stays measured in their original units~\cite{dragicevic2012my}. We took those percent differences (one per participant per metric) and analyzed them in the following ways.\\

\begin{figure}
    \centering
    \includegraphics[width=\columnwidth]{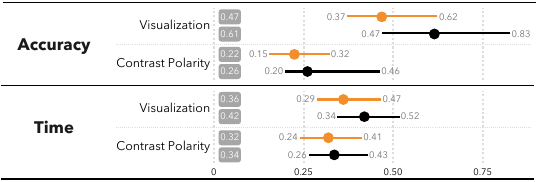}
    \caption{
        Bootstrapped average (\meanicon) percent differences for YA~\sqboxblack{yaOrange} and PLA~\sqboxblack{plaBlack} resulting from each design factor (contrast polarity, visualization) and for each metric (accuracy, time). Confidence intervals are Bonferroni-corrected for 4 comparisons.
    }
    \label{fig:perc-diff-ci}
\end{figure}

\noindent\textbf{\textit{Does the Impact of Contrast Polarity Shift with Age?}}
First, to measure whether the individual impacts of contrast polarity change with age, we calculated the average and standard error (SE) of percent differences for both time and accuracy for each age group. This entailed aggregating a group's percent differences for one metric into a single vector and generating 10000 bootstrapped samples (with replacement) that were subsequently used for calculating the mean and a 95\% confidence interval (CI) for each design. 
We compared the impact of contrast polarity on each age group using interval analysis~\cite{cumming2014new}. This approach is a non-parametric method in which the percentage of CI overlap is used to characterize the strength of differences between two groups. An interval overlap percentage (IOP) less than or equal to $0$ is interpreted as \textit{strong} evidence of a difference, while $0 < IOP < 0.5$ is categorized as \textit{weak} evidence and $IOP\geq 0.5$ is deemed \textit{insufficient} evidence. As a non-parametric method that does not rely on assumptions of normality or equal variance, IOP provides a more nuanced interpretation of data differences, categorizing them into meaningful levels of evidence compared to the binary output (significant, not significant) of methods such as the $t$-test and ANOVA~\cite{besancon2017difference}. Furthermore, we used Bonferroni-corrected intervals to minimize the possibility of Type-I errors due to multiple comparisons, widening the CIs and reducing the probability of observing strong evidence of differences when several conditions are compared~\cite{armstrong2014use}. These results are labelled \textsc{Contrast Polarity} in \autoref{fig:perc-diff-ci} and reported in \autoref{sec:acc-results} and \autoref{sec:time-results}.\\

\noindent\textbf{\textit{Is Contrast Polarity as Impactful as Visualization?}}
To contextualize the impact of contrast polarity as a \textit{non-data encoding}, we compared its percent differences with those caused by visualization type, a \textit{data encoding} known to affect performance~\cite{davis2022risks,saket2018task}. Because we used three visualizations, we averaged the percent differences between each pair and calculated bootstrapped confidence intervals (\autoref{fig:perc-diff-ci}, \textsc{Visualization}) for comparison. Visualization type had a noticeably greater impact on accuracy for both age groups. However, the distributions of percent differences for time were similar for both visualization and polarity across age groups, indicating comparable impacts on performance.\\

\begin{figure}
    \centering
    \includegraphics[width=\columnwidth]{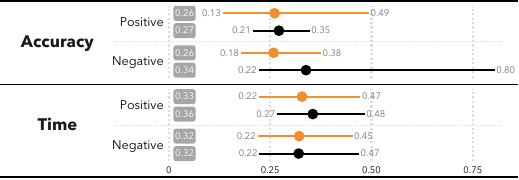}
    \caption{
        Bootstrapped average (\meanicon) percent differences for YA~\sqboxblack{yaOrange} and PLA~\sqboxblack{plaBlack}, with participants grouped by their better-performing contrast polarity for each metric (accuracy, time). Confidence intervals are Bonferroni-corrected for 4 comparisons. 
    }
    \label{fig:perc-diff-cp-comp}
\end{figure}

\noindent\textbf{\textit{Does One Contrast Polarity Have a Greater Impact?}}
To understand whether either contrast polarity led to more dramatic changes in performance (\iDotE larger percent differences), we calculated bootstrapped confidence intervals (CIs) for accuracy and time after splitting the data per age group into those who performed better with each contrast polarity, shown in \autoref{fig:perc-diff-cp-comp}. When comparing across both polarities and age groups, the confidence intervals in the figure almost completely overlap, indicating that both polarities can lead to similar performance improvements regardless of age.\\

\begin{figure}
    \centering
    \includegraphics[width=\columnwidth]{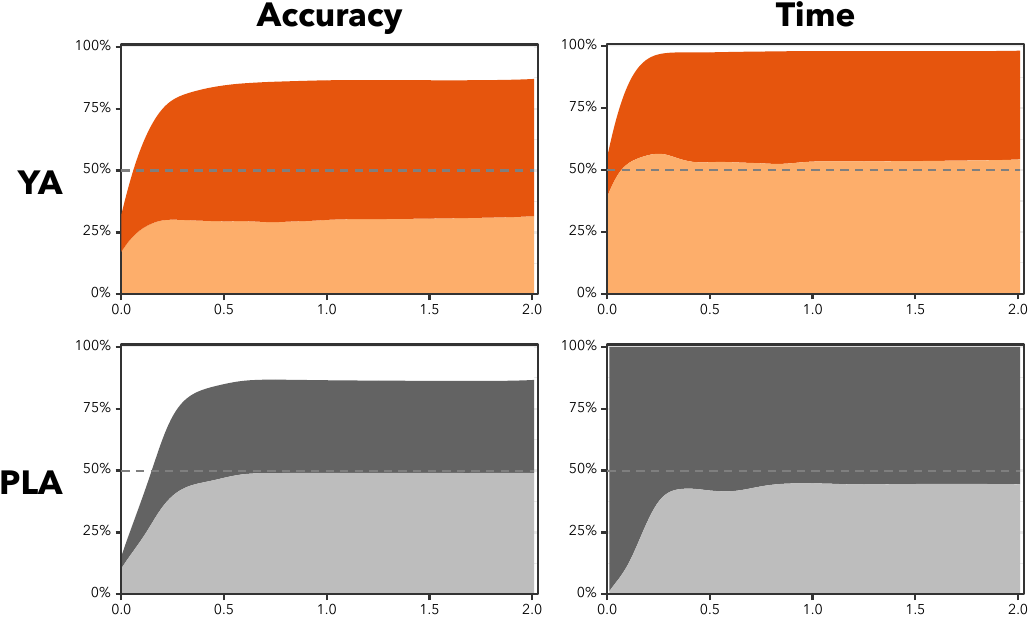}
    \caption{Proportions of participants who performed best with positive (YA~\sqboxblack{propLightOrange}, PLA~\sqboxblack{propLightGray}) and negative contrast polarity (YA~\sqboxblack{propDarkOrange}, PLA~\sqboxblack{propDarkGray}), and those who performed equally well with both~\sqboxblack{white}. Participants are filtered across increasing maximum threshold values of percent difference ($y$-axis), with data grouped by metric~(accuracy, time) and age group. A dashed line at $y=50\%$ is provided for reference.}
    \label{fig:con-pol-prop}
\end{figure}

\noindent\textbf{\textit{Is One Contrast Polarity More Frequently Advantageous?}}
To determine if 
each contrast polarity benefited similar amounts of people, 
we calculated the ratio of participants with higher average performance for each polarity, categorized by age group and metric (accuracy or time) and depicted as a proportion plot for varying percent differences in \autoref{fig:con-pol-prop}. 
A greater number of YA had their best accuracy with negative contrast, while a similar proportion of PLA had their best accuracy with positive contrast. However, for time, the distribution was roughly equal for both age groups. 
This suggests that each age group has a different contrast polarity more often suited for greater accuracy, while similar proportions of each group can achieve greater speed with either polarity.\\ 

\begin{figure}
    \centering
    \includegraphics[width=\columnwidth]{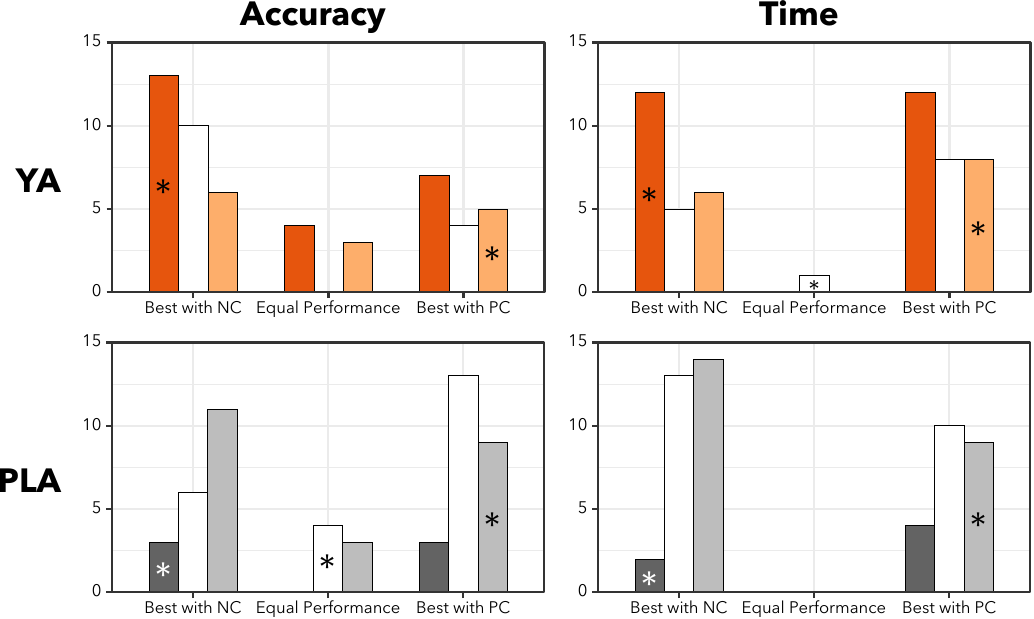}
    \caption{
    The number of YA who \textbf{preferred} each contrast polarity (positive~\sqboxblack{propLightOrange}, negative~\sqboxblack{propDarkOrange}, or no preference~\sqboxblack{white}) as well as PLA who \textbf{preferred} each contrast polarity (positive~\sqboxblack{propLightGray}, negative~\sqboxblack{propDarkGray}, or no preference~\sqboxblack{white}), grouped by the contrast polarity (negative = NC, positive = PC, equal performance) with which they achieved their \textbf{best performance} (accuracy, time). 
    Bars where the preferred polarity matches the best-performing polarity are marked with an \textbf{\textasteriskcentered}.
    }
    \label{fig:pref-vs-best}
\end{figure}

\noindent\textbf{\textit{Do People Prefer Their Best-Performing Contrast Polarity?}}
To investigate whether participants preferred the contrast polarity that they performed best with, we also analyzed how participants' preferred contrast polarity (as per the pre-study questionnaire) compared with their empirically-best contrast polarity (higher average accuracy or lower average time), shown in \autoref{fig:pref-vs-best}. Overall, we noticed disagreement between the two, indicating that what participants prefer may not actually be what they are best-suited to use, perhaps due to general familiarity or other outside factors~\cite{sethi2023dark}. 

\subsection{Accuracy}
\label{sec:acc-results}
The percent differences in accuracy caused by contrast polarity for PLA ranged from $0$ to $2.0$ ($\mu = 0.26$), while YA's ranged from $0$ to $0.96$ ($\mu = 0.22$).
Approximately $92.3\%$ of PLA and $88.4\%$ of YA had a percent difference less than or equal to $0.5$. Approximately half of PLA had a percent difference less than or equal to $0.22$, while about half of YA were less than or equal to $0.18$. The IOP analysis between confidence intervals of percent differences for PLA and YA found insufficient evidence of differences ($IOP \geq 0.5$), indicating that the strength of contrast polarity's effect on accuracy does not noticeably change with age. CIs for percent differences for participants who had higher accuracy with each of the contrast polarities showed insufficient evidence of differences ($IOP \geq 0.5$) both between contrast polarities and between age groups, indicating that 
both contrast polarities can lead to similar improvements, and that this trend is not impacted by aging.

Similar proportions of PLA achieved higher performance with each contrast polarity (48.1\% positive, 38.5\% negative, 13.5\% equal), while there was a slightly larger disparity for YA (30.7\% positive, 55.8\% negative, 13.5\% equal). Compared to time, we observed a larger portion of participants in both age groups achieving equal performance between contrast polarities, \iDotE a percent difference of $0$, which may be a result of time measurements having a larger range of values than accuracy based on limited trials.
The interval analysis found strong evidence of differences $(IOP < 0)$ between percent differences in accuracy based on contrast polarity and visualization in both age groups, indicating that participants generally experienced larger differences in accuracy due to the choice of visualization. This indicates that visualization choice has a more pronounced impact on accuracy than contrast polarity, which stays consistent in late adulthood.

\subsection{Time}
\label{sec:time-results}
The percent differences in time due to contrast polarity for PLA ranged from $0.01$ to $1.12$ ($\mu = 0.34)$, while YA's ranged from $0.0$ to $0.93$ ($\mu = 0.32$). Around $80.8\%$ of PLA had a percent difference less than or equal to $0.5$ compared to $73.1\%$ of YA. Approximately half of PLA and YA had percent differences less than or equal to $0.29$ and $0.23$, respectively. The interval analysis comparing percent differences in time between YA and PLA based on contrast polarity found insufficient evidence of differences ($IOP \geq 0.5$), suggesting that the strength of contrast polarity's impact on visual analysis time does not noticeably change with age. We further observed insufficient evidence of differences in percent differences for each polarity, both between and within age groups. This indicates that both contrast polarities can lead to similar time improvements, and that this trend is consistent with increasing age. 

Compared to accuracy, the proportions of participants achieving better response times with each contrast polarity were much closer, with $44.2\%$ of PLA and $53.8\%$ of YA excelling with positive contrast whereas $55.8\%$ and $44.2\%$, respectively, excelled with negative contrast. Noticeably fewer participants had equal performance with both polarities, with no PLA and only 1 younger adult ($1.9\%$) achieving a percent difference of $0$. 
The interval analysis found insufficient evidence $(IOP \geq 0.5)$ in percent differences between visualization type and contrast polarity for both age groups. This suggests that contrast polarity is roughly as impactful on performance time as visualization type, regardless of age.

\section{Discussion and Future Work}
Our results indicate that, for both YA and PLA, negative and positive contrast polarity are similarly capable of improving user performance. 
Moreover, contrast polarity can be just as impactful as visualization choice when time is an important design consideration. Thus, we recommend that practitioners offer both positive and negative contrast color palettes of visualizations when possible, perhaps as a universally-helpful personalization feature. 

Our results differ from those reported by Piepenbrock et al.~\cite{piepenbrock2013positive}, which observed better reading performance using positive contrast polarity; our findings, on the other hand, point to \textit{both} contrast polarities supporting similar improvements in accuracy and time (in similar proportions) for visual analysis tasks. However, both our study and their study observed results that were consistent across YA and PLA. The disparities between the two studies' findings in terms of a ``better'' contrast polarity may indicate a difference in how it impacts reading versus how it impacts visual analysis.

Our findings regarding preference versus performance differ from Saket et al.~\cite{saket2018task}, who observed that user preferences for visualization type correlated with visualizations that improved performance. This might be because contrast polarity preferences, unlike those for visualization type, are impacted by broader aspects of the user experience such as phone menus and app designs. Since this preference is likely decided outside of a visualization context, it may not correspond with visualization performance. Additionally, users may not realize how contrast polarity, a \textit{non-data encoding} only affecting color, is affecting their performance, whereas the immediate impact of the choice of visualization type, a \textit{data-encoding}, on performance is more noticeable (\eDotG trying to use a Pie chart to assess correlation) and is thus more aligned with preferences.

These results also illustrate the importance of carefully choosing data analysis methods. Participants in our study experienced visualization stimuli with both positive and negative contrast, allowing the possibility for some to achieve noticeably higher performance with one than the other; simply calculating averages and performing interval analysis to find differences led to the misleading initial conclusion that contrast polarity had \textit{no} effect on performance. Analyzing the data at an individual level and focusing on changes in performance illustrated the nuanced, varied impacts that each contrast polarity can have, consistent with recent work cautioning against design recommendations based on the ``average observer''~\cite{davis2022risks}.

Future work can investigate the impact of other non-data-encoding design choices, such as graph stroke width and shape, graph orientation, and gridlines, and existing accessibility technologies, such as screen readers, on the performance of YA and PLA, as well as how their impact compares to those of data encodings. 

\section{Conclusion}
This work presents the results of a crowdsourced study investigating the impact of contrast polarity on the graphical analysis performance of younger adults as well as people in late adulthood. We found that positive and negative contrast polarities can each positively impact performance (and in equal measure) across both age groups. Furthermore, we observed strong evidence that contrast polarity has an impact on response times similar to the choice of visualization, suggesting its worthiness as a customization feature. 


\balance
\def\UrlBreaks{\do\/\do-}
\bibliographystyle{abbrv-doi}
\bibliography{bibliography.bib}

\begin{thebibliography}{10}

\bibitem{ibm-carbon-design}
Color palettes.
\newblock \url{https://carbondesignsystem.com/data-visualization/color-palettes/}.
\newblock (Accessed on 02/17/2022).

\bibitem{apple-visual-design}
Dark mode.
\newblock \url{https://developer.apple.com/design/human-interface-guidelines/ios/visual-design/dark-mode/}.
\newblock (Accessed on 02/19/2022).

\bibitem{google-material-design}
Dark theme.
\newblock \url{https://material.io/design/color/dark-theme.html}.
\newblock (Accessed on 02/18/2022).

\bibitem{Developi34:online}
Developing websites for older people: How web content accessibility guidelines (wcag) 2.0 applies | web accessibility initiative (wai) | w3c.
\newblock \url{https://www.w3.org/WAI/older-users/developing/}.
\newblock (Accessed on 03/22/2022).

\bibitem{Prolific80:online}
Prolific | online participant recruitment for surveys and market research.
\newblock \url{https://www.prolific.co/}.
\newblock (Accessed on 06/24/2021).

\bibitem{wcag2023}
Web content accessibility guidelines (wcag) 2.2.
\newblock \url{https://www.w3.org/TR/WCAG22/}, October 2023.

\bibitem{alexander2011passive}
G.~L. Alexander, B.~J. Wakefield, M.~Rantz, M.~Skubic, M.~A. Aud, S.~Erdelez, and S.~Al~Ghenaimi.
\newblock Passive sensor technology interface to assess elder activity in independent living.
\newblock {\em Nursing research}, 60(5):318--325, 2011.

\bibitem{andrew2024light}
S.~Andrew, C.~Bishop, and G.~W. Tigwell.
\newblock Light and dark mode: A comparison between android and ios app ui modes and interviews with app designers and developers.
\newblock {\em Proceedings of the ACM on Interactive, Mobile, Wearable and Ubiquitous Technologies}, 8(1):1--23, 2024.

\bibitem{armstrong2014use}
R.~A. Armstrong.
\newblock When to use the b onferroni correction.
\newblock {\em Ophthalmic and Physiological Optics}, 34(5):502--508, 2014.

\bibitem{berk2022development}
L.~E. Berk.
\newblock {\em Development through the lifespan}.
\newblock Sage Publications, 2022.

\bibitem{besancon2017difference}
L.~Besan{\c c}on and P.~Dragicevic.
\newblock {La Diff{\'e}rence Significative entre Valeurs p et Intervalles de Confiance}.
\newblock In AFIHM, ed., {\em {29{\`e}me conf{\'e}rence francophone sur l'Interaction Homme-Machine}}, p.~10. {AFIHM}, Poitiers, France, 2017.
\newblock Alt.IHM.

\bibitem{chung2009contrast}
S.~T. Chung and J.~S. Mansfield.
\newblock Contrast polarity differences reduce crowding but do not benefit reading performance in peripheral vision.
\newblock {\em Vision Research}, 49(23):2782--2789, 2009.

\bibitem{cumming2014new}
G.~Cumming.
\newblock The new statistics: Why and how.
\newblock {\em Psychological science}, 25(1):7--29, 2014.

\bibitem{davis2022risks}
R.~Davis, X.~Pu, Y.~Ding, B.~D. Hall, K.~Bonilla, M.~Feng, M.~Kay, and L.~Harrison.
\newblock The risks of ranking: Revisiting graphical perception to model individual differences in visualization performance.
\newblock {\em IEEE Transactions on Visualization and Computer Graphics}, 2022.

\bibitem{dragicevic2012my}
P.~Dragicevic.
\newblock {\em My technique is 20\% faster: Problems with reports of speed improvements in HCI}.
\newblock PhD thesis, Inria Saclay Ile de France, 2012.

\bibitem{johnson2017designing}
J.~Johnson and K.~Finn.
\newblock {\em Designing user interfaces for an aging population: Towards universal design}.
\newblock Morgan Kaufmann, 2017.

\bibitem{morey-heart-failure}
S.~A. Morey, L.~H. Barg-Walkow, and W.~A. Rogers.
\newblock Managing heart failure on the go: Usability issues with mhealth apps for older adults.
\newblock {\em Proceedings of the Human Factors and Ergonomics Society Annual Meeting}, 61(1):1--5, 2017. doi: {{%
10\hspace{.1pt}\discretionary{.}{%
}{.}\hspace{.4pt}1177\discretionary{/}{%
}{/}1541931213601496}}


\bibitem{pearson1999data}
R.~K. Pearson.
\newblock Data cleaning for dynamic modeling and control.
\newblock In {\em 1999 European Control Conference (ECC)}, pp. 2584--2589. IEEE, 1999.

\bibitem{piepenbrock2014positive}
C.~Piepenbrock, S.~Mayr, and A.~Buchner.
\newblock Positive display polarity is particularly advantageous for small character sizes: implications for display design.
\newblock {\em Human Factors}, 56(5):942--951, 2014.

\bibitem{piepenbrock2013positive}
C.~Piepenbrock, S.~Mayr, I.~Mund, and A.~Buchner.
\newblock Positive display polarity is advantageous for both younger and older adults.
\newblock {\em Ergonomics}, 56(7):1116--1124, 2013.

\bibitem{rubin1989psychophysics}
G.~S. Rubin and G.~E. Legge.
\newblock Psychophysics of reading. vi—the role of contrast in low vision.
\newblock {\em Vision Research}, 29(1):79--91, 1989.

\bibitem{saket2018task}
B.~Saket, A.~Endert, and {\c{C}}.~Demiralp.
\newblock Task-based effectiveness of basic visualizations.
\newblock {\em IEEE Transactions on Visualization and Computer Graphics}, 25(7):2505--2512, 2018.

\bibitem{sethi2023dark}
T.~Sethi and M.~Ziat.
\newblock Dark mode vogue: Do light-on-dark displays have measurable benefits to users?
\newblock {\em Ergonomics}, 66(12):1814--1828, 2023.

\bibitem{spehar2002role}
B.~Spehar.
\newblock The role of contrast polarity in perceptual closure.
\newblock {\em Vision Research}, 42(3):343--350, 2002.

\bibitem{westheimer2003visual}
G.~Westheimer, P.~Chu, W.~Huang, T.~Tran, R.~Dister, et~al.
\newblock Visual acuity with reversed-contrast charts: Ii. clinical investigation.
\newblock {\em Optometry and vision science}, 80(11):749--752, 2003.

\bibitem{while2024glanceable}
Z.~While, T.~Blascheck, Y.~Gong, P.~Isenberg, and A.~Sarvghad.
\newblock Glanceable data visualizations for older adults: Establishing thresholds and examining disparities between age groups.
\newblock {\em arXiv preprint arXiv:2403.12343}, 2024.

\bibitem{while2024gerontovis}
Z.~While, R.~J. Crouser, and A.~Sarvghad.
\newblock Gerontovis: Data visualization at the confluence of aging.
\newblock {\em arXiv preprint arXiv:2403.13173}, 2024.

\bibitem{whitlock}
L.~A. Whitlock and A.~C. McLaughlin.
\newblock Identifying usability problems of blood glucose tracking apps for older adult users.
\newblock {\em Proceedings of the Human Factors and Ergonomics Society Annual Meeting}, 56(1):115--119, 2012. doi: {{%
10\hspace{.1pt}\discretionary{.}{%
}{.}\hspace{.4pt}1177\discretionary{/}{%
}{/}1071181312561001}}


\end{thebibliography}

\end{document}